\documentclass[11pt]{article}
\usepackage{amsfonts%,ulem
}
\usepackage{amsmath}
\usepackage{amssymb}
\usepackage{graphicx}
\usepackage{color}
\usepackage{braket}
\usepackage[utf8]{inputenc}
\usepackage{hyperref} 

\def \be {\begin{equation}}
\def \ee {\end{equation}}
\def \ba {\begin{aligned}}
\def \ea {\end{aligned}}
\def \bea {\begin{eqnarray}}
\def \eea {\end{eqnarray}}

\def \fa {\partial_\theta \phi}
\def \ga {\partial_\theta \bar\phi}
\def \Epa {E^+_{\theta}}
\def \Ema {E^-_{\theta}}
\def \Ept {E^+_{t}}
\def \Emt {E^-_{t}}
\def \ft {\partial_t \phi}
\def \gt {\partial_t \bar\phi}
\def \cS {\mathcal{S}}

\def \qt {\partial_t \varphi}
\def \qa {\partial_\theta \varphi}

\def \At {A_t}
\def \Ax {A_\theta}
\def \Bt {\bar A_t}
\def \Bx {\bar A_\theta}

\begin{document}

\begin{titlepage}
\begin{flushright}
NORDITA 2020-058 \\
June,  2020
\end{flushright}
\vspace{0.5cm}
\begin{center}
{\Large \bf $T\bar{T}$ deformation of chiral bosons and Chern-Simons AdS$_3$ gravity}
\lineskip .75em
\vskip 2.5cm
{Hao Ouyang and Hongfei Shu }
\vskip 2.5em
 {\normalsize\it 
Nordita, KTH Royal Institute of Technology and Stockholm University\\
Roslagstullsbacken 23, SE-106 91 Stockholm, Sweden
}
\vskip 3.0em
\end{center}
\begin{abstract}
We study the $T\bar{T}$ deformation of the chiral bosons and show the equivalence between the chiral bosons of opposite chiralities and the scalar fields at the Hamiltonian level under the deformation. We also derive the deformed Lagrangian of more generic theories which contain an arbitrary number of chiral bosons to all orders. By using these results, we derive the  $T\bar{T}$ deformed boundary action of the AdS$_3$ gravity theory in the Chern-Simons formulation. We compute the deformed one-loop torus partition function, which satisfies the $T\bar{T}$ flow equation up to the one-loop order. Finally, we calculate the deformed stress–energy tensor of a solution describing a BTZ black hole in the boundary theory, which coincides with the boundary stress–energy tensor derived from the BTZ black hole with a finite cutoff. 
  \end{abstract}
\end{titlepage}

\baselineskip=0.7cm

%\tableofcontents
%\newpage

\section{Introduction}
The deformation by the $T\bar{T}$ operator \cite{Zamolodchikov:2004ce} has drawn much attention, because of its solvability and the relation with gravity theory. Although the $T\bar{T}$ deformation is an irrelevant deformation, it is possible to derive the deformed Lagrangian, finite size spectrum and the S-matrix from the ones of the original theory \cite{Caselle:2013dra,Smirnov:2016lqw, Cavaglia:2016oda}, which does not require the integrability in many cases. Based on the finite size spectrum, one could compute the torus partition function of the $T\bar{T}$ deformation \cite{Cardy:2018sdv, Datta:2018thy,Aharony:2018bad}, which is still modular invariant but not conformal invariant. The $T\bar{T}$ deformation is related to the gravity theory in several aspects. On the one hand, the deformed theory can be interpreted as the original theory coupled to a topological gravity \cite{Dubovsky:2017cnj,Dubovsky:2018bmo,Cardy:2018sdv}. More concretely, one finds a one-to-one map between the equations of motion (EOM) in the deformed theories and those of the original theories \cite{Conti:2018tca, Conti:2019dxg}, which enables one to derive the all-order deformed Lagrangians \cite{Coleman:2019dvf}. On the other hand, the two-dimensional $T\bar{T}$ deformed holographic CFT is proposed to correspond to the gravity theory with a finite cutoff under the Dirichlet boundary condition, where the cutoff is explicitly related to the deformation parameter \cite{McGough:2016lol}.  More discussions on the holography under the $T\bar{T}$ deformation can be found in \cite{Kraus:2018xrn,Cottrell:2018skz,Taylor:2018xcy,Hartman:2018tkw,Caputa:2019pam,Guica:2019nzm,Gross:2019ach}. See also \cite{Jiang:2019hxb} for an interesting review and related topics.

 The $T\bar{T}$ deformation of the two-dimensional scalar theory has been well studied. In particular, the all-order deformed action of the $N$ massless free bosons has the form of the Nambu-Goto action in the static gauge of $N+2$-dimension \cite{Cavaglia:2016oda}. The deformation of a scalar with an arbitrary potential was shown in \cite{Cavaglia:2016oda,tateocdd} and more examples of Lagragians of $T\bar{T}$ deformed theories was presented in \cite{Bonelli:2018kik}.
 In this paper, we study the chiral bosons, which are interesting in many aspects such as string theory and condensed matter. Even though the chiral bosons are not manifestly Lorentz invariant, the sum of a left and a right chiral bosons reproduces the scalar theory \cite{Sonnenschein:1988ug,Tseytlin:1990ar}. The $T\bar{T}$ deformed action of a general system of chiral bosons,  scalars and fermions was studied in \cite{Frolov:2019nrr}, where the first-order action for chiral bosons and canonical stress–energy tensor were used. We are interested here in the Floreanini-Jackiwaction action \cite{Floreanini:1987as} and the covariant stress–energy tensor.
 
 A remarkable connection between chiral Wess-Zumino-Witten (WZW) models and Chern-Simons theories was established in \cite{Witten:1988hf,Moore:1989yh,Elitzur:1989nr}. 
 In particular, the AdS$_3$ Einstein gravity theory can be reformulated as a $SL(2,\mathbb{R})\times SL(2,\mathbb{R})$ Chern-Simons theory \cite{Witten:1988hc}. Much attention has been paid to the exact boundary action \cite{Coussaert:1995zp, Witten:2007kt,Barnich:2017jgw,Cotler:2018zff}, due to its connection with two-dimensional conformal field theory \cite{Brown:1986nw}.
 The AdS$_3$ Chern-Simons action can be reduced to two chiral $SL(2,\mathbb{R})$ Wess-Zumino-Witten (WZW) models on the boundary,
and the AdS$_3$ boundary condition implements certain constraints on the chiral WZW model \cite{Coussaert:1995zp}. In \cite{Cotler:2018zff}, the exact boundary action was shown to be a quantum field theory of reparametrizations, which analogous to the Schwarzian action of the nearly AdS$_2$ gravity. Moreover, the torus partition is one-loop exact and shows a shift in the central charge of $13$. 
The present work aims to study the $T\bar{T}$ deformation of the 
boundary action. We derive the $T\bar{T}$ deformed Lagrangian of generic chiral boson theories by explicitly solving the flow equation. Then we focus on the $T\bar{T}$ deformed action of the constrained chiral WZW model associated with the AdS$_3$ Chern-Simons theory. We will see how the exact boundary action and the one-loop torus partition function change under the $T\bar{T}$ deformation. We also calculate the deformed stress–energy tensor of the boundary theory for the BTZ black hole, and compare it with the boundary stress–energy tensor derived from the BTZ black hole with a finite cutoff. Our results provide a concrete realization of the $T\bar{T}$ deformation on the boundary of the Chern-Simons AdS$_3$ gravity and may shed light on the holography dual of the deformation.

This paper is organized as follows. In section 2, we present the $T\bar{T}$ deformed Lagrangian of chiral boson theories. We show the equivalence between the sum of two chiral bosons of opposite chiralities and a massless free non-chiral scalar under the $T\bar{T}$ deformation at the Hamiltonian level. In section 3, we review the relation between the AdS$_3$ Chern-Simons theory and the sum of two constrained $SL(2,\mathbb{R})$ chiral WZW model  of opposite chiralities, and derive the corresponding deformed Lagrangian. We then compute the one-loop torus $T\bar{T}$ deformed partition function, which is found to satisfy the flow equations of $T\bar{T}$ deformation in all-order of deformation parameter up to one-loop level. We also compute the deformed stress–energy tensor for a solution describing a BTZ black hole in the deformed field theory and compare it with the boundary stress–energy tensor of the BTZ black hole at a finite cutoff. Section 4 is devoted to conclusions and discussions. In Appendix \ref{A}, we consider the $J\bar J$ and $T\bar J$ deformation of the chiral bosons. In Appendix \ref{B}, we study the solutions to the EOMs of $T\bar{T}$ deformed WZW models.

\section{$T \bar T$ deformed Lagrangian of chiral bosons}
In this section, we will study the $T\bar{T}$ deformation of chiral bosons.
The Floreanini-Jackiw action \cite{Floreanini:1987as} of a left-moving chiral boson is
\begin{equation}
S_{\mathrm{left}}=\int d^2x\frac{1}{2}(\ft\fa- \fa \fa ).
\end{equation}
As a warm-up, we will first consider the  simple case of two chiral bosons of opposite chiralities and solve the flow equation induced by the $T \bar T$ deformation.
More complicated theories of chiral bosons will also be considered, which will be useful in the study of the AdS$_3$ Chern-Simons theory.

\subsection{Two chiral bosons of opposite chiralities}
Let us begin with the undeformed Lagrangian of a left and a right chiral boson
\begin{align}
S_0=&\int d^2x \mathcal{L}_0,\\
\mathcal{L}_0=&-\frac{1}{2} \left( -\ft\fa+\frac{\Ept}{\Epa} \fa \fa 
+\gt\ga-\frac{\Emt}{\Ema} \ga \ga \right) ,\label{lsimple}
\end{align}
where $E^a$ is the zweibein and the metric is $g_{\mu\nu}=E^+_\mu E^-_\nu+E^-_\mu E^+_\nu$. We couple the zweibein to the fields such that the undeformed action is invariant under the transformation
\begin{equation}
\begin{split}
\delta \phi=&\epsilon_+^\theta\fa,\label{eq:tran}\\
\delta \bar \phi=&\epsilon_-^\theta \ga,\\
\delta E^+_\mu=&\epsilon_+^\theta \partial_{\theta}  E^+_\mu
+ E^+_\theta    \partial_{\mu}       \epsilon_+^{\theta}.\\
\delta E^-_\mu=&\epsilon_-^\theta \partial_{\theta}  E^-_\mu
+ E^-_\theta    \partial_{\mu}       \epsilon_-^{\theta},
\end{split}
\end{equation}
where $\epsilon_\pm^\theta$ are coordinate dependent transformation parameters. The translation symmetry generated by constant  $\epsilon_\pm^\theta$ enables us to define the stress–energy tensor as
\begin{equation}\label{tij}
T^{\mu}_{~\nu}
=-\frac{1}{\det E}\frac{\delta S}{\delta E^A_\mu} E^A_\nu.
\end{equation} 
When $E^A_\mu$ are constants the conserved law can be written as $\partial_\mu T^{\mu}_{~\nu}=0$.

The $T\bar T$ deformation of a two-dimensional field theory is induced by the $T\bar{T}$ operator which is defined as minus the determinant of the stress–energy tensor.
Concretely, the $T\bar T$ deformed Lagrangian $\mathcal{L}_\lambda$ is the solution to the flow equation
\begin{equation} \label{eq:flow}
\frac{\partial \mathcal{L}_\lambda}{\partial \lambda}=\det E\det T_\lambda,
\end{equation}
with the initial condition $(\ref{lsimple})$.
Here $\lambda$ is the deformation parameter and 
$T_\lambda$ is the deformed stress–energy tensor of the deformed theory.
We can solve the equation by making a perturbative expansion in small $\lambda$  and then guessing the exact solution. 
Skipping the boring details, the solution is given by
\begin{equation}
\label{ltt}
\mathcal{L}_\lambda=\frac{1}{2}(\ft \fa-\gt \ga )
-\frac{(\Ema \Ept+\Emt\Epa)(\fa \fa-\ga\ga)}{4\Epa \Ema}+\frac{\det E}{2{\lambda}}(\cS-1),
\end{equation}
with
\begin{equation}
\cS=\sqrt{1-\frac{(\fa \fa+\ga \ga)}{\Ema \Epa}\lambda+\frac{(\fa \fa-\ga\ga)^2}{4\Ema {}^2 \Epa{}^2}\lambda^2}.
\end{equation}
The deformed theory still has the conservation laws $\partial_\mu (T^{\mu}_{~\nu})_{{\lambda }}=0$, which corresponds to the symmetries
\begin{align}
\delta \phi=&\epsilon_1\fa
+\epsilon_0 \fa \frac{2 \Epa \Ema-\lambda(\fa \fa-\ga \ga)}{2\Epa \Ema \cS},\\
\delta \bar \phi=&\epsilon_1 \ga
-\epsilon_0 \ga \frac{2 \Epa \Ema+\lambda(\fa \fa-\ga \ga)}{2\Epa \Ema \cS},
\end{align}
where $\epsilon_i$ are constant parameters. We also consider the $J\bar J$ and $T\bar J$ deformation and the results are shown in Appendix \ref{A}.

\subsection{Equivalence to the $T\bar{T}$ deformation of a  non-chiral free scalar}

In the undeformed theory, the sum of a left moving chiral boson and a right moving chiral boson is equivalent to a free massless scalar \cite{Sonnenschein:1988ug,Henneaux:1987hz}. We now show that the equivalence still holds under $T \bar T$ deformation.

We now restrict our attention to flat spacetime so we can set $\Ept=\Epa=\Ema=-\Emt=1/\sqrt{2}$ after solving the flow equation. Therefore the Lagrangian (\ref{ltt}) becomes
\begin{equation}\label{ltt0}
\mathcal{L}_\lambda=\frac{1}{2}(\ft \fa-\gt \ga)+\frac{1}{{2\lambda}}(\cS-1),
\end{equation}
with
\begin{equation}
\cS=\sqrt{1-2{(\fa \fa+\ga \ga)}\lambda+(\fa \fa-\ga\ga)^2\lambda^2},
\end{equation}
The stress–energy tensor of the deformed theory becomes 
\begin{align}\label{tmn}
(T_\lambda)^{\mu}_{~\nu}=\left(\begin{array}{cc}
\frac{1-\cS}{2\lambda}& \frac{(\fa \fa-\ga\ga)}{2} \\ 
-\frac{(\fa \fa-\ga\ga)}{2}& \frac{\cS -1}{2\lambda \cS}+\frac{(\fa \fa-\ga\ga)^2\lambda}{2 \cS}
\end{array} \right),
\end{align}
from which we obtain the corresponding energy and momentum
\begin{align}
H_{\lambda}&=\int d\theta \frac{1}{2\lambda}(1-\sqrt{1-2{(\fa \fa+\ga \ga)}\lambda+{(\fa \fa-\ga\ga)^2}\lambda^2}),\label{Hc}\\
P_{\lambda}&=\int d\theta \frac{1}{2}(\fa \fa-\ga \ga).\label{Pc}
\end{align}

The Hamiltonian density of the system can be written as
\begin{equation}\label{Hchiral}
\mathcal{H}=\frac{1}{2\lambda}(1-\sqrt{1-8{(\pi^2+\bar\pi ^2)}\lambda
	+16{(\pi^2-\bar\pi^2)^2}\lambda^2}),
\end{equation}
where $\pi=\frac{1}{2} \fa$ and $\bar\pi=-\frac{1}{2} \ga$ are the canonical momenta of the fields.

Let us turn to the $T \bar T$ deformed  free massless  non-chiral scalar. The Lagrangian is given by \cite{Cavaglia:2016oda}
\begin{equation}
\mathcal{L}^\mathrm{scalar}_{\lambda}=\frac{1}{2\lambda}(\sqrt{1+2\lambda(\qt \qt-\qa \qa)}-1).
\end{equation}
The associated Hamiltonian density is
\begin{equation}\label{HLo}
\mathcal{H}^\mathrm{scalar}_{\lambda}=\frac{1}{2\lambda}(1-\sqrt{(1-2 \lambda \qa^2)(1-2 \lambda \pi_{\varphi}^2 )}).
\end{equation}
where the  canonical moment of $\varphi$ is defined as
\begin{equation}
\pi_{\varphi}=\frac{\qt}{\sqrt{1+2\lambda(\qt \qt-\qa \qa)}}.
\end{equation}
One can check
that the Hamiltonian densities (\ref{Hchiral}) and (\ref{HLo}) are equivalent via the relation
\begin{equation}
\varphi=\frac{1}{\sqrt{2}} (\phi+\bar \phi),~~~
\pi_{\varphi}={\sqrt{2}} (\pi+\bar \pi).
\end{equation}

The $T \bar T$ deformed Lorentz {invariant} free massless scalar  is related to the undeformed model via a field dependent coordinate transformation\cite{Conti:2018tca,Conti:2019dxg}. 
To obtain a solution to the deformed theory,  one can start with a solution 
\begin{equation}
\varphi(\tilde t, \tilde\theta)=f(\tilde{x}^+)+g(\tilde{x}^-),
\end{equation}
where $\tilde{x}^\pm= \tilde\theta \pm \tilde t$,
to the equation of motion of the undeformed model
\begin{equation}
(\partial^2_{\tilde t}-
\partial^2_{\tilde\theta}) \varphi =0.
\end{equation}
Then one need to solve the equations 
\begin{equation}\label{co}
t=\tilde t+\frac{\lambda}{2}\big(G(\tilde x^-)-F(\tilde x^+)\big),~~~ 
\theta=\tilde\theta+\frac{\lambda}{2}\big(G(\tilde x^-)+F(\tilde x^+)\big),
\end{equation}
to express $\tilde x^\pm$ in terms of $t$ and $\theta$, where the derivatives of $F$ and $G$ are the components of  stress–energy tensor in the specific classical solution
\begin{equation}
F'(x)=2f'(x)f'(x),~~~G'(x)=2g'(x)g'(x).
\end{equation}
A solution to the  equation of motion of the deformed model is then given by
\begin{equation}\label{sl}
\varphi(t, \theta)=f\big(\tilde x^+{(t,\theta)}\big)+g\big(\tilde x^-{(t,\theta)}\big).
\end{equation}

Let us return to the chiral boson model. Though we cannot find a coordinate transformation which maps
the equations of motion of the model (\ref{ltt0})  directly  to those of the undeformed model, one can check that
\begin{equation}\label{sc}
\phi(t,\theta)=h(t)+\sqrt{2}f(\tilde x^+(t,\theta)),~~~\bar\phi(t,\theta)=\bar h(t)+\sqrt{2}g(\tilde  x^-(t,\theta)),
\end{equation}
is a solution to the equations of motion. Here $t, \theta$ are still related to $\tilde x^\pm$ by (\ref{co}). $h(t)$ and $\bar{h}(t)$ are arbitrary functions of $t$. 
We show this in a more general case in Appendix \ref{B}.
The  energy and momentum corresponding to the solution (\ref{sc}) are 
\begin{align}
H_{\lambda}&=\int d\theta \frac{f'(\tilde x^+)^2-g'(\tilde x^-)^2-4\lambda f'(\tilde x^+)^2 g'(\tilde x^-)^2}{1-4 \lambda^2 f'(\tilde x^+)^2 g'(\tilde x^-)^2},\\
P_{\lambda}&=\int d\theta \frac{f'(\tilde x^+)^2-g'(\tilde x^-)^2}{1-4 \lambda^2 f'(\tilde x^+)^2 g'(\tilde x^-)^2}.
\end{align}

We now put the deformed model on a circle of length $L$. Then the fields should be periodic in coordinate $\theta$.
We take periodicities of $f$ and $g$ to be  $L$ and consider the solutions with the following form
\begin{equation}\label{sold}
\varphi=f(n(\lambda)\tilde x^+(t,\theta))+g(m(\lambda)\tilde  x^-(t,\theta)),
\end{equation}
where we introduce $n(\lambda)$ and $m(\lambda)$ such that the periodicity of $\varphi$ is $L$ in coordinate $\theta$. It is not difficult to show that 
\begin{equation}
\tilde{t}(t,L)-\tilde{t}(t,0)=\lambda P_\lambda,~~~
\tilde{x}(t,L)-\tilde{x}(t,0)=-\lambda H_\lambda.
\end{equation}
Then we have
\begin{equation}
n(\lambda)=\frac{L}{L-\lambda(H_\lambda- P_\lambda)},~~~
m(\lambda)=\frac{L}{L-\lambda(H_\lambda+ P_\lambda)}.
\end{equation}
Using equations (\ref{co}) and $F(L)-F(0)=H_0+P_0$, $G(L)-G(0)=H_0-P_0$, we get
\begin{equation}
-H_0 L+H_\lambda L\pm P_0 L\mp P_\lambda L-\lambda H_\lambda^2+\lambda P_\lambda^2=0.
\end{equation}
Finally we get
\begin{equation}
H_\lambda=\frac{L-\sqrt{L^2-4 H_0 L \lambda +4 P_0^2 \lambda^2}}{2 \lambda},~~~ P_\lambda=P_0,
\end{equation}
which is a classical version of
the general quantum spectrum in \cite{Smirnov:2016lqw,Cavaglia:2016oda}.

\subsection{General theory of chiral bosons}
To solve the flow equation, the field contents and details of the potentials are not important. We can study the $T\bar{T}$ deformed Lagrangian of more general model of chiral bosons with the initial translational invariant Lagrangian
\begin{equation}
\mathcal{L}_0=
C-\frac{\Ept}{2\Epa} K_++\frac{\Emt}{2\Ema}  K_- +\Epa V_+ +\Ema V_-+\Ept W_+ +\Emt W_-,
\end{equation}
where $K_\pm$, $W_\pm$ and $V_\pm$ are the functions of the fields. We require that the equations of motion are consistent with the conservation of the stress–energy tensor defined by (\ref{tij}). We can again solve the flow equation (\ref{eq:flow}) using a perturbative approach. The all-order solution can be written as
   \begin{equation}
   \begin{split}
   \mathcal{L}_\lambda=&C
   +\frac{\tilde E^-_{\theta} \tilde E^+_{t}+\tilde E^-_{t} \tilde E^+_{\theta}}{4\tilde E^+_{\theta} \tilde E^-_{\theta}}(K_--K_+)
   +\frac{1}{2\lambda}(\frac{\Epa \Ema }{\tilde E^+_{\theta} \tilde E^-_{\theta}}\det \tilde E \mathcal{S}+\det \tilde E-2 \det E).
   \end{split}
   \end{equation}
   with
   \begin{align}
   \tilde E^\pm_{t}&= E^\pm_{t}\mp \lambda  V_{\mp}, %\\
   \tilde E^\pm_{\theta}%&
   = E^\pm_{\theta}\mp \lambda  W_{\mp}, \\
   \mathcal{S}&=\sqrt{\frac{4\tilde E^{+2}_{\theta} \tilde E^{-2}_{\theta}
   -4(K_-+K_-)\tilde E^+_{\theta} \tilde E^-_{\theta}\lambda
    +(K_--K_-)^2 \lambda^2	}{4\Epa \Ema}}.
    \end{align}

As a particular example, we consider a generalized chiral bosons theory
\begin{align}
\mathcal{L}_0=& \frac{1}{2} \left( F_{t\theta}-\frac{\Ept}{\Epa} F_{\theta\theta} 
-\bar F_{t\theta}+\frac{\Emt}{\Ema} \bar F_{\theta\theta}\right) -\Epa V(\phi) -\Ema \bar V(\bar\phi) ,\label{gen}\\
F_{\mu\nu}=&G^{IJ}(\phi)\partial_{\mu}\phi_I\partial_{\nu}\phi_J,~~~
I,J=1,2,...,N,\\
\bar F_{\mu\nu}=&\bar G^{\bar I \bar J}(\bar \phi)\partial_{\mu}\bar\phi_{\bar I}\partial_{\nu}\bar\phi_{\bar J},~~~
\bar I,\bar J=1,2,...,\bar N,
\end{align}
where $G$ and $\bar G$ are non-degenerate matrices. 
The $T\bar{T}$ deformation of the chiral boson theory (\ref{gen}) is therefore
	\begin{equation}
	\begin{split}\label{lttg}
	\mathcal{L}_\lambda=&\frac{1}{2}(F_{t\theta}-\bar F_{t\theta} )-\Ema \bar V-\Epa V
	-\frac{(\Ema \Ept+\Emt\Epa+\lambda (\Epa V-\Ema \bar V))(F_{\theta\theta}-\bar F_{\theta\theta})}{4\Epa \Ema}\\
	&+\frac{\det E-\lambda(\Ema \bar V+\Epa V)}{2{\lambda}}(\cS-1),
		\end{split}
	\end{equation}
with
	\begin{equation}
	\cS=\sqrt{1-\frac{(F_{\theta\theta}+\bar F_{\theta\theta})}{\Ema \Epa}\lambda+\frac{(F_{\theta\theta}-\bar F_{\theta\theta})^2}{4\Ema {}^2 \Epa{}^2}\lambda^2}.
	\end{equation}

One can also add an arbitrary
number of  Weyl-Majorana fermions to the theory (\ref{gen}).
The undeformed Lagrangian is
\begin{equation}
\begin{split}
    \mathcal{L}_0=& \frac{1}{2} \left( F_{t\theta}-\frac{\Ept}{\Epa} F_{\theta\theta} 
-\bar F_{t\theta}+\frac{\Emt}{\Ema} \bar F_{\theta\theta}\right) -\Epa V(\phi) -\Ema \bar V(\bar\phi)    \\
&+ B_{MN}\psi^M(\Ept\partial_\theta-\Epa\partial_t)
\psi^N
-\bar B_{MN}\bar\psi^M(\Emt\partial_\theta-\Ema\partial_t)
\bar\psi^N,
\end{split}
\end{equation}
where $\psi$ and $\bar \psi$ are Weyl-Majorana left and right fermions respectively. The deformed theory is then given by (\ref{lttg}) with
\begin{equation}
\begin{split}
C=&\frac{1}{2} \left( F_{t\theta}
-\bar F_{t\theta}\right) ,~~~
K_+= F_{\theta\theta} ,~~~K_-= \bar F_{\theta\theta},\\
V_+=&V(\phi)+ B_{MN}\psi^M\partial_t\psi^N,~~~
V_-=\bar V(\phi)- \bar B_{MN}\bar \psi^M\partial_t\bar \psi^N,\\
W_+=& B_{MN}\psi^M\partial_\theta\psi^N,~~~
W_-=- \bar B_{MN}\bar \psi^M\partial_\theta\bar \psi^N.
\end{split}   
\end{equation}
This action differs from the one obtained in \cite{Frolov:2019nrr} using the canonical stress–energy tensor.
It was argued in \cite{Frolov:2019nrr} that there should be a  field redefinition which would make the $T\bar T$ deformed action driven by the canonical stress–energy tensor conicide with the one  driven by the covariant stress–energy tensor. It would be interesting to find such a field redefinition explicitly.

\section{$T\bar{T}$ deformation and Chern-Simons gravity}

The three-dimensional Einstein gravity theory with a negative cosmological constant can be reformulated as a  Chern-Simons action with a gauge group $SL(2,\mathbb{R})\times SL(2,\mathbb{R})$ \cite{Witten:1988hc}. It was shown in \cite{Coussaert:1995zp} that the Chern-Simons action  is equivalent to two copies of constrained $SL(2,\mathbb{R})$ chiral WZW models of opposite chiralities on the boundary, which  can be combined into a non-chiral Liouville field theory.
The chiral description is more convenient to deal with the zero modes and leads to
geometric actions associated with coadjoint orbits of the Virasoro group\cite{Barnich:2017jgw,Cotler:2018zff}.
In this section, we will focus on the $T\bar{T}$ deformed action of the constrained chiral WZW model. Since the original Lagrangian is a special case of (\ref{gen}), we can get the all-order $T\bar{T}$ deformation of the boundary action  using the results in the previous section \footnote{See also \cite{Leoni:2020rof} for the $T\bar{T}$-deformation of the classical Liouville field theory.}.

\subsection{AdS$_3$ Chern-Simons theory}\label{sec:CS-AdS3}
Let us recall the connection between AdS$_3$ gravity and the chiral WZW model derived in \cite{Coussaert:1995zp}.
The AdS$_3$ Einstein gravity with metric 
\be
\label{eq:AdS-ds}
ds^{2}=-(r^{2}+1)dt^{2}+r^{2}d\theta^{2}+\frac{dr^{2}}{r^{2}+1},
\ee
can be reformulated as the Chern-Simons action \cite{Cotler:2018zff}
\be
\ba
\label{eq:CS}
S&=S[A]-S[\bar{A}]+S_{{\rm bdy}},\\
S[A]&=-\frac{k}{2\pi}\int_{{\cal M}}dt\wedge{\rm Tr}\Big(-\frac{1}{2}\tilde{A}\wedge\dot{\tilde{A}}+A_{0}\tilde{F}\Big),\\
S_{{\rm bdy}}&=-\frac{k}{4\pi}\int_{\partial{\cal M}}dx^{2}\Big(\frac{\Ept}{\Epa}{\rm Tr}(A_{\theta}^{2})-\frac{\Ept}{\Epa}{\rm Tr}(\bar{A}_{\theta}^{2})\Big),
\ea
\ee
where $k=\frac{1}{4G}$ and we couple the boundary terms to the boundary zweibein $E^a$.
We will take $\Ept=\Epa=\Ema=-\Emt=1/\sqrt{2}$.
The gauge fields $A$ and $\bar{A}$ are expressed by using the $SL(2)$ generators and related with the bulk dreibein $e^a$ and the bulk spin connection $\omega$
\be
A-\bar{A}=2e,\quad A+\bar{A}=2\omega.
\ee
In this action, $A=A_0dt+\tilde{A}_idx^i$ and $\bar{A}=\bar{A}_0dt+\tilde{\bar{A}}_idx^i$ are separated into the temporal and spatial parts. The boundary conditions of the gauge fields are fixed to be $A_{-}=A_{t}-A_{\theta}=0$ and  $\bar{A}_{-}=\bar{A}_{t}+\bar{A}_{\theta}=0$, which are chosen to match the asymptotics of the AdS$_3$ geometry
\be
\label{eq:bdy-cond}
\begin{split}
A=&
\left(\begin{array}{cc}
\frac{1}{2}\Omega+\frac{dr}{2r}+O(r^{-2}) & O(r^{-1})\\
\sqrt{2}rE^+dx^{+}+O(r^{-1}) & -\frac{1}{2}\Omega-\frac{dr}{2r}+O(r^{-2})
\end{array}\right), \\
\bar{A}=&\left(\begin{array}{cc}
\frac{1}{2}\Omega-\frac{dr}{2r}+O(r^{-2}) & -\sqrt{2}rE^-dx^{-}+O(r^{-1})\\
O(r^{-1}) & -\frac{1}{2}\Omega+\frac{dr}{2r}+O(r^{-2})
\end{array}\right),
\end{split}
\ee
 where $\Omega$ is the boundary spin connection. For simplicity, we consider $\Omega=0$ in this paper. 
The boundary term $S_{\rm bdy}$ is necessary for a consistency variation principle. 

Since the spatial field strength $\tilde{F}$ is flat, one can parametrize the $\tilde{A}$ and $\tilde{\bar{A}}$ as
\be
\tilde{A}=g^{-1}\tilde{d}g,\quad\tilde{\bar{A}}=\bar{g}^{-1}\tilde{d}\bar{g},
\ee
where $\tilde{d}$ is the spatial exterior derivative. $g$ and $\bar{g}$ are elements of $SL(2)$ and can be written in the Gauss parameterization:
\be
\ba
g&=\left(\begin{array}{cc}
1&0 \\ 
F&1
\end{array} \right)
\left(\begin{array}{cc}
e^\varphi&0 \\ 
0&e^{-\varphi}
\end{array} \right)
\left(\begin{array}{cc}
1&\Psi \\ 
0&1
\end{array} \right),\\
\bar g&=\left(\begin{array}{cc}
1&- \bar F\\ 
0&1
\end{array} \right)
\left(\begin{array}{cc}
e^{-\bar\varphi}&0 \\ 
0&e^{\bar\varphi}
\end{array} \right)
\left(\begin{array}{cc}
1& 0\\ 
-\bar\Psi &1
\end{array} \right).
\ea
\ee
The gauge fields can be written as
\be
\ba
\tilde{A}&=g^{-1}dg=\left(\begin{array}{cc}
A^3&A^- \\ 
A^+ &-A^3
\end{array} \right)
=\left(\begin{array}{cc}
-e^{2\varphi}\Psi dF+d\varphi&-e^{2\varphi}\Psi^2 dF+2\Psi d\varphi + d\Psi\\ 
e^{2\varphi} dF &e^{2\varphi}\Psi dF-d\varphi
\end{array} \right),\\
\tilde{\bar A}&=\bar g^{-1}d \bar g=\left(\begin{array}{cc}
\bar A^3&\bar A^- \\ 
\bar A^+ &-\bar A^3
\end{array} \right)
=\left(\begin{array}{cc}
e^{2\bar\varphi}\bar\Psi d \bar F-d \bar\varphi&-e^{2\bar\varphi} d\bar F\\ 
e^{2\bar\varphi}\bar\Psi^2 d \bar F-2\bar\Psi d\bar\varphi - d{\bar{\Psi}} &-e^{2\bar\varphi}\bar\Psi d \bar F+d \bar\varphi
\end{array} \right).
\ea
\ee
The action (\ref{eq:CS}) thus can be evaluated as
\be
\label{eq:ac-WZW}
S=\frac{k}{\pi}\int_{\partial {\cal M}}d^2x{\cal L}^{\rm WZW}_0,
\ee
where ${\cal L}^{\rm WZW}_0$ has the form of equation (\ref{gen}) with $\Ept=\Epa=\Ema=-\Emt=1/\sqrt{2}$ and
\begin{equation}\label{eq:WZW2}
\begin{split}
&F_{t\theta}=\partial_t \varphi \partial_\theta \varphi +e^{2\varphi}\partial_\theta F \partial_t \Psi,~~~
\bar F_{t\theta}=\partial_t \bar\varphi \partial_\theta \bar\varphi +e^{2\bar\varphi}\partial_\theta \bar F \partial_t \bar\Psi,\\
&F_{\theta\theta}=A^3_\theta A^3_\theta +A^+_\theta A^-_\theta,~~~
\bar F_{\theta\theta}=\bar A^3_\theta \bar A^3_\theta+\bar  A^+_\theta \bar  A^-_\theta,~~~
V=\bar V=0.
\end{split}
\end{equation}
The fields in the expression of $g$ and $\bar{g}$ are not independent.
The boundary condition (\ref{eq:bdy-cond}) imposes the constrains
\begin{equation}
     A^3_\theta=\bar A^3_\theta=0,~~~
     A^+_\theta-\sqrt{2} E^+_\theta r=0,~~~ 
     \bar A^-_\theta+\sqrt{2} E^-_\theta r=0
\end{equation}
on the AdS boundary.
The constrains can be expressed as
\be
\ba
\label{eq:bdy-cond-gauss}
e^{\varphi}&=\sqrt{\frac{r}{\partial_{\theta}F}},\quad\Psi=-\frac{\partial_{\theta}^{2}F}{2r\partial_{\theta}F},\\e^{\bar{\varphi}}&=\sqrt{\frac{r}{\partial_{\theta}\bar{F}}},\quad\bar{\Psi}=-\frac{\partial_{\theta}^{2}\bar{F}}{2r\partial_{\theta}\bar{F}}.
\ea
\ee
By using the conditions (\ref{eq:bdy-cond-gauss}), we could express the action $S$ in terms of $F$ and $\bar{F}$, which we parameterize as
\begin{equation}
\label{eq:F-phi}
F=\tan \frac{\phi}{2},~~~\bar F=\tan \frac{\bar \phi}{2}.
\end{equation}
where $\phi$ and $\bar\phi$ are elements of  ${\rm Diff}(S^1)/{ PSL}(2,\mathbb{R})$ and we get two copies of the Alekseev-Shatashvili quantization of  coadjoint orbit ${\rm Diff}(S^1)/{ PSL}(2,\mathbb{R})$ of the Virasoro group \cite{Alekseev:1988ce}.
If we parameterize $F$ and $\bar{F}$ as
\begin{equation}
\label{eq:F-phi2}
F=\tan \frac{\alpha\phi}{2},~~~\bar F=\tan \frac{\alpha \bar \phi}{2},
\end{equation}
with $\alpha\neq n$, $n\in \mathbb{Z}$, we get the orbit ${\rm Diff}(S^1)/U(1)$. See \cite{Witten:1987ty} for further discussion.

\subsection{$T\bar{T}$ deformation of the boundary action}

With the solution  (\ref{lttg}) to the flow equation induced by the $T\bar{T}$ deformation at hand,  we are now ready to get the all-order $T\bar{T}$ deformation of the boundary action. Simply plugging (\ref{eq:WZW2}) into  (\ref{lttg}), we obtain a $T\bar{T}$ deformed WZW model denoted by
$\mathcal{L}_\lambda^{\mathrm{WZW}}$
\footnote{In principle $\lambda$ should be rescaled to keep the flow equation invariant  due to the coefficient $k/\pi$ in front of the action. But for convenience we are not going to rescale $\lambda$ here }.
However, the action (\ref{eq:ac-WZW}) is  constrained. It is a non-trivial question whether the constrains are deformed by the $T\bar{T}$.
To treat the constraints carefully, we
introduce Lagrange multipliers in the undeformed action:
\begin{equation}
\mathcal{L}_0^{\mathrm{cWZW}}=\mathcal{L}_0^{\mathrm{WZW}}
-a_3 A^3_\theta-\bar a_3 \bar A^3_\theta
-a_+ (A^+_\theta-\sqrt{2} E^+_\theta r)-\bar a_-( \bar A^-_\theta+\sqrt{2} E^-_\theta r).
\end{equation}
Here we keep $E^\pm_\theta$ unfixed which is necessary when we apply the solution (\ref{lttg}).
The terms with coefficients $E^\pm_\theta$ can be view as potentials.
By using (\ref{lttg}) again, we find the deformed constrained Lagrangian
\begin{equation}
\begin{split}
\mathcal{L}_\lambda^{\mathrm{cWZW}}=&\mathcal{L}_\lambda^{\mathrm{WZW}}
-a_3 A^3_\theta-\bar a_3 \bar A^3_\theta-a_+ (A^+_\theta-\frac{r+r\mathcal{S}}{2}-\frac{r\lambda( F_{\theta\theta}-\bar F_{\theta\theta})}{2})\\
&
-\bar a_-( \bar A^-_\theta+\frac{r+r\mathcal{S}}{2}-\frac{r\lambda( F_{\theta\theta}-\bar F_{\theta\theta})}{2}).
\end{split}
\end{equation}
The constraints $A^3_\theta=\bar A^3_\theta=0$ are solved by
\begin{equation}\label{con2}
\Psi=\frac{e^{-2\varphi}\partial_\theta\varphi}{\partial_\theta F},~~~
\bar\Psi=\frac{e^{-2\bar\varphi}\partial_\theta\bar\varphi}{\partial_\theta\bar F}.
\end{equation}
Plugging $A^3_\theta=\bar A^3_\theta=0$ into the rest two constraints, we get
\begin{equation}\label{ed6}
 A^+_\theta=r+r^2\lambda\bar A^+_\theta,~~~\bar A^-_\theta=-r+r^2\lambda  A^-_\theta,
\end{equation}
which are similar to (5.16) in \cite{Llabres:2019jtx}.
The explicit expressions are
\be
\ba
\label{eq:bdy-ps}
&p-1-\frac{\lambda}{4{\bar{p}}^3}(4 \bar p^2 \bar s+3 {(}\partial_\theta \bar p{)}^2-2\bar p \partial_\theta^2\bar  p )=0,\\
&\bar p-1-\frac{\lambda}{4p^3}(4 p^2 s+3 {(}\partial_\theta p{)}^2-2p \partial_\theta^2 p )=0,
\ea
\ee
where
\begin{equation}
e^{2\varphi}=\frac{r p}{\partial_\theta F},~~~
e^{2\bar \varphi}=\frac{r\bar  p}{\partial_\theta\bar F},~~~
s=\frac{1}{2}\{ F,\theta \},~~~
\bar s=\frac{1}{2}\{\bar F,\theta \}.
\end{equation}
$s$ and $\bar{s}$ are halves of the Schwarzian derivatives defined by
\begin{equation}
\{f,\theta \}=\frac{\partial_\theta^3 f}{ \partial_\theta f}-\frac{3}{2}\left(\frac{\partial_\theta^2 f}{\partial_\theta f}\right)^2.
\end{equation}
Suppose the solution $p$ and $\bar{p}$ satisfying (\ref{eq:bdy-ps}) is known, we then substitute this solution to the ${\cal L}^{\rm cWZW}_\lambda$ and obtain the all-order $T\bar{T}$ deformed Lagrangian
\be
\ba
\mathcal{L}_\lambda^{\mathrm{cWZW}}=& \frac{s}{2p}+ \frac{\bar s}{2 \bar p} -
	\frac{\dot F''}{4F'}+\frac{3\dot F' F''}{8F'^2}
	+\frac{\dot {\bar F}''}{4\bar F'}-\frac{3\dot {\bar F}' \bar F''}{8 \bar F'^2}+\frac{3 \left(p'\right)^2}{8 p^3}-\frac{p''}{4 p^2}\\
	&
	-\frac{3 \dot{p} p'}{8 p^2}+\frac{\dot{p}'}{4 p}
	+\frac{3 \left(\bar{p}'\right)^2}{8 \bar{p}^3}-\frac{\bar{p}''}{4 \bar{p}^2}
	+\frac{3 \dot{\bar{p}} \bar{p}'}{8 \bar{p}^2}-\frac{\dot{\bar{p}}'}{4 \bar{p}},
\ea
\ee
where the overdot and prime denote the derivative with respect to $t$ and $\theta$ respectively. The deformed  stress–energy tensor is given by
\begin{equation}
(T_\lambda)^{\mu}_{~\nu}=
\frac{k}{\pi}\left(
\begin{array}{cc}
 -\frac{p+\bar{p}-2}{2 \lambda } & \frac{p-\bar{p}}{2 \lambda } \\
 -\frac{p-\bar{p}}{2 \lambda } & \frac{p^2-2 \bar{p} p+p+\bar{p}^2+\bar{p}-2}{2 \lambda  \left(p+\bar{p}-1\right)} \\
\end{array}
\right).
\end{equation}

Using the  parameterization
\begin{equation}
F=\tan \frac{\alpha \phi}{2},~~~\bar F=\tan \frac{\alpha\bar \phi}{2}.
\end{equation}
We have
\begin{equation}
s=\frac{\phi^{(3)}}{2\phi'}-\frac{3\phi''^2}{4\phi'^2}+\frac{\alpha\phi'^2}{4},~~~
\bar s=\frac{\bar \phi^{(3)}}{2\bar \phi'}-\frac{3\bar \phi''^2}{4\bar \phi'^2}+\frac{\alpha\bar \phi'^2}{4},
\end{equation}
where $f^{(n)}$ denotes the $n$th derivative of $f$ with respect to $\theta$.
We also define
\begin{equation}
u=\frac{\dot \phi''}{2\phi'}-\frac{3\dot\phi' \phi''}{4\phi'^2}+\frac{\alpha\dot\phi \phi'}{4},~~~
\bar u=\frac{\dot {\bar \phi}''}{2\bar \phi'}-\frac{3\dot {\bar \phi}' \bar \phi''}{4\bar \phi'^2}+\frac{\alpha\dot{\bar  \phi} \bar \phi'}{4}.
\end{equation}
Dropping total derivatives, the Lagrangian can be written as  
\begin{equation}
\begin{split}
\mathcal{L}=&\frac{s }{2p}+\frac{\bar s }{2 \bar p}-\frac{u }{2}+\frac{\bar u }{2}
+\frac{3 \left(p'\right)^2}{8 p^3}-\frac{p''}{4 p^2}
-\frac{3 \dot{p} p'}{8 p^2}+\frac{\dot{p}'}{4 p}
	+\frac{3 \left(\bar{p}'\right)^2}{8 \bar{p}^3}-\frac{\bar{p}''}{4 \bar{p}^2}
	+\frac{3 \dot{\bar{p}} \bar{p}'}{8 \bar{p}^2}-\frac{\dot{\bar{p}}'}{4 \bar{p}},
\end{split}
\end{equation}
where $p$ and $\bar p$ are determined by $s$ and $\bar s$ through the constraints  (\ref{eq:bdy-ps}).

%Let us now  solve the constraints (\ref{eq:bdy-ps}) to get the deformed Lagrangian. 
Though the constraints (\ref{eq:bdy-ps}) are difficult to solve to all orders in $\lambda$, we can solve $p$ and $\bar p$ in the first few orders of small $\lambda$
\begin{align}
p=&1+\lambda \bar s+\lambda^2(-s\bar s-\frac{ s''}{2})+O(\lambda )^3, \label{eq:cons-ps1}\\
\bar p=&1+\lambda s+\lambda^2(-s\bar s-\frac{ \bar s''}{2})+O(\lambda )^3,\label{eq:cons-ps2}
\end{align}
which leads to
\begin{equation}
\begin{split}
\mathcal{L}_\lambda^{\mathrm{cWZW}}
=& \frac{s}{2}+ \frac{\bar s}{2 } -
\frac{\dot F''}{2F'}+\frac{3\dot F' F''}{4F'^2}
+\frac{\dot {\bar F}''}{2\bar F'}-\frac{3\dot {\bar F}' \bar F''}{4 \bar F'^2}-\lambda  s \bar{s}\\
&+
\frac{1}{8} \lambda ^2 \Big(4 \bar{s} s''+8 s^2 \bar{s}+8 s' \bar{s}'+3 \bar{s}'^2-3 \dot{\bar{s}} \bar{s}'-2 \bar{s} \dot{\bar{s}}'+4 s \bar{s}''+6 \bar{s} \bar{s}''+8 s \bar{s}^2\\
&
+6 s s''+3 s'^2+3 \dot{s} s'+2 s \dot{s}'\Big)+O\left(\lambda ^3\right).
\end{split}
\end{equation}
When $\lambda=0$, this reproduces the original Lagrangian.  The first order term $s\bar{s}$ is nothing but the $T\bar{T}$ operator of the undeformed action.

At the end of this subsection, let us comment on the deformed constrains (\ref{ed6}) and their relation with finite cutoff AdS. Since (\ref{eq:bdy-cond-gauss}) is derived from the boundary condition (\ref{eq:bdy-cond}) of gauge fields (or metric), it is natural to guess that the boundary condition will also be transformed non-trivially under the $T\bar{T}$ deformation. Let us suppose the new boundary is at $r=r_c$ with a large enough $r_c$. If we identify $r_c^2 \lambda=1$, (\ref{ed6}) leads to $2e^{\pm}_{\theta}=r_c$, which is consistency with the metric (\ref{eq:AdS-ds}) at finite cutoff $r=r_c$. In section \ref{sec:CS-BTZ}, we will check the identification $r_c^2 \lambda=1$ in more details by calculating the boundary stress–energy tensor.

\subsection{One-loop torus partition function}
The partition function in the undeformed theory was obtained and shown to be  one-loop exact in \cite{Cotler:2018zff}. We now compute the one-loop torus partition function in the deformed theory.
Let us Wick-rotate to the Euclidean time $t=-iy$ and put the boundary theory on a torus of complex structure $\tau$. The Euclidean action is $S_E=-i S$. On the torus, the coordinate $z=\theta+i y$ has the identifications $z\sim z+2\pi$ and $z\sim z+2\pi \tau$. We first focus on the  ${\rm Diff}(S^1)/{ PSL}(2,\mathbb{R})$ case. The fields $\phi$ and $\bar{\phi}$ satisfy the boundary condition
\be
\ba
f(\theta+2\pi,y)&=f(\theta,y)+2\pi,\quad
 f(\theta,y)=f(\theta+2\pi{\rm Re}(\tau),y+2\pi{\rm Im}(\tau)).
\ea
\ee
We consider the saddle point of the Euclidean Lagrangian
\begin{equation}
\phi_0=\bar\phi_0=\theta-\frac{ \tau_1}{\tau_2}y,~~~
p_0=\bar p_0=\frac{1}{2\gamma },
\end{equation}
where
$\tau_1$ 
and $\tau_2$ are real and imaginary part
of $\tau$ respectively and we will use 
\begin{equation}
\gamma=\frac{\sqrt{\lambda +1}-1}{\lambda },
\end{equation}
instead of $\lambda$ to avoid square root.

Expanding $\phi$ and $\bar \phi$ in fluctuations around the saddle
\begin{equation}
\phi=\phi_0+\delta\phi,~~~\bar\phi=\bar\phi_0+\delta\bar\phi,
\end{equation}
the fluctuations of $p$ and $\bar p$ depend on  $\delta\phi$ and $\delta\bar \phi$ via the constraints.
We have
\begin{equation}
p=p_0+p_1+p_2+...,~~~
\bar p=\bar p_0+\bar p_1+\bar p_2+...
\end{equation}
where $p_1$ ($\bar p_1$) and $p_2$ ($\bar p_2$) are linear and quadratic  terms in the fluctuation fields $\delta\phi$ and $\delta\bar \phi$ respectively. We then expand the Lagrangian and the constraints around the saddle, and express every term by using $\delta\phi$ and $\delta\bar \phi$. On the torus, the fluctuation fields  $\delta\phi$ and $\delta\bar \phi$ can be expand as
\begin{equation}
\delta\phi=\sum_{\substack{m,n \\ n\neq -1,0,1}}\epsilon_{m,n}f_{m,n},~~~\delta\bar \phi=\sum_{\substack{m,n \\ n\neq -1,0,1}}\bar\epsilon_{m,n}f_{m,n},
\end{equation}
where we have set the zero modes to zero and the functions
\be
f_{m,n}=\frac{1}{2\pi}\exp\left(i\frac{m y}{\tau_2}+i n(\theta -\frac{\tau_1}{\tau_2}y)\right),
\ee
satisfy
\be
\int_{T^2} d^2x f_{m_1,n_1}f_{m_2,n_2}=\tau_2\delta_{m_1,-m_2}\delta_{n_1,-n_2}.
\ee
Then $p_1$ and $\bar p_1$ are solved by
\begin{equation}
\begin{split}
p_1=&\sum_{\substack{m,n \\ n\neq -1,0,1}}q_n\big((1-2 \gamma ) (2 n^2-1)\epsilon_{m,n}+ \bar\epsilon_{m,n}\big)f_{m,n},\\
\bar p_1=&\sum_{\substack{m,n \\ n\neq -1,0,1}}q_n\big((1-2 \gamma ) (2 n^2-1)\bar\epsilon_{m,n}+ \epsilon_{m,n}\big)f_{m,n}.    
\end{split}
\end{equation}
where
\begin{equation}
q_n=-\frac{i (2 \gamma -1) n \left(n^2-1\right)}{2  \left(-2 \gamma  n^2+\gamma+n^2-1\right) \left(-2 \gamma  n^2+\gamma+n^2\right)}.
\end{equation}
Finally, the  quadratic action is given by
\begin{equation}\label{qa1}
-\frac{\pi}{k}S_E=\frac{\gamma\tau_2}{2}(2\pi)^2+\sum_{\substack{m,n \\ n\neq -1,0,1}}
(\epsilon_{m,n}, \bar\epsilon_{m,n})M_{m,n}
\left(\begin{array}{c}
\epsilon_{-m,-n} \\ 
\bar\epsilon_{-m,-n}
\end{array} \right),
\end{equation}
where $M_{m,n}$ is a $2\times2$ matrix
\begin{equation}
\begin{split}
M_{m,n}=& \frac{ n (n^2-1)}{16 \left(n^2 (\chi -1)+1\right)^2 \left(\chi -n^2 (\chi -1)\right)^2}
\left(
\begin{array}{cc}
A_{m,n}(\tau_1,\tau_2) &B_{m,n}(\tau_1,\tau_2) \\
B_{m,n}(\tau_1,\tau_2) &-A_{m,n}(\tau_1,-\tau_2) \\
\end{array}
\right),
\end{split}
\end{equation}
with $\chi=\frac{\gamma}{1-\gamma}=\frac{1}{\sqrt{\lambda +1}}$ and
\begin{align}
A_{m,n}(\tau_1,\tau_2)=&-2 i \chi  \left(n^2 \left(n^2-1\right) (\chi
-1)^2-\chi \right) \left(m+i n \tau _2 \chi -n \tau
_1\right)\nonumber\\
&-n \left(n^2-1\right)^2 \tau _2 \chi  \left(\chi
^2-1\right)^2,\\
B_{m,n}(\tau_1,\tau_2)=&-n \left(n^2-1\right) \tau _2 \chi  \left(\chi
^2-1\right) \left(n^2 (\chi -1)^2-\chi ^2-1\right).
\end{align}
The determinant of $M_{m,n}$ is
\begin{equation}
\det M_{m,n}=\frac{n^2 \left(n^2-1\right)^2 \chi ^2 \left(m-i n \tau
	_2 \chi -n \tau _1\right) \left(m+i n \tau _2 \chi
	-n \tau _1\right)}{64 \left(n^2 (\chi -1)+1\right)^2
	\left(\chi -n^2 (\chi -1)\right)^2}.
\end{equation} 
Then following the procedure in \cite{Cotler:2018zff}, we obtain the classical partition function \begin{equation}
Z_{\mathrm{c}}=\exp \left(2\pi C \tau_2 \frac{  \sqrt{\lambda +1}-1}{6 \lambda }\right),~~~C=6k.
\end{equation}
and the one-loop torus partition 
\begin{equation}
\begin{split}
Z_{1-\mathrm{loop}}=&\exp \left(2\pi\tau_2\big( \frac{ C  \left(\sqrt{\lambda +1}-1\right)}{6 \lambda }+\frac{13}{12\sqrt{\lambda+1}}\big)\right)\\
&\times\left|\prod_{n=2}^{\infty}\frac{1}{1-\exp(2\pi i n(\tau_1+i \frac{\tau_2}{\sqrt{\lambda+1}}))}\right|^2.    
\end{split}
\end{equation}
Note that the partition function is not modular invariant even in the undeformed theory.
It is easy to check that the classical partition satisfies the flow equation on the torus \cite{Aharony:2018bad}
 \begin{equation}
 \label{eq:flow-eq-cla}
 -\frac{\pi C}{6}\partial_{\lambda}Z_{\mathrm{c}}=(\frac{\tau_2}{4}( \partial_{\tau_2}^2+\partial_{\tau_1}^2)
 +\frac{\lambda }{2}( \partial_{\tau_2} -\tau_2^{-1})\partial_{\lambda})Z_{\mathrm{c}},
 \end{equation}
 while the one-loop partition function satisfies the flow equation up to the one-loop
 \begin{equation}
 \label{eq:flow-eq-loop}
-\frac{1}{ Z_{\mathrm{c}}}\frac{\pi  }{6}\partial_{\lambda}Z_{1-\mathrm{loop}}
=\frac{1}{C Z_{\mathrm{c}}}(\frac{\tau_2}{4}( \partial_{\tau_2}^2+\partial_{\tau_1}^2)
+\frac{\lambda }{2}( \partial_{\tau_2} -\tau_2^{-1})\partial_{\lambda})Z_{1-\mathrm{loop}}+O(C^{-1}),
\end{equation}
where the first term on the right hand side is order $O(C)$. The one-loop torus function satisfy the flow equation up to the one loop order $O(C^0)$. This suggests that the $T\bar{T}$ deformed partition function should not be one-loop exact as in the undeformed theory. It is worth to note that the flow equations (\ref{eq:flow-eq-cla}) and (\ref{eq:flow-eq-loop}) are satisfied for all order in  $\lambda$, which provide evidence for our all-order $T\bar{T}$ deformed Lagrangian.

One can also compute the partition function of the Diff($\mathbb{S}^1$)/$U(1)$ case, where
\begin{equation}
F=\tan \frac{\alpha \phi}{2},~~~\bar F=\tan \frac{\alpha \bar \phi}{2},
\end{equation}
with $\alpha\neq n$, $n\in \mathbb{Z}$.
One can repeat the same steps and finally get
\begin{equation}
\begin{split}
Z_{1-\mathrm{loop}}=&\exp \left(2\pi\tau_2\big( \frac{ C  \left(\sqrt{\alpha^2\lambda +1}-1\right)}{6 \lambda }+\frac{13}{12\sqrt{\alpha^2\lambda+1}}\big)\right)\\
&\times\left|\prod_{n=1}^{\infty}\frac{1}{1-\exp(2\pi i n(\tau_1+i \frac{\tau_2}{\sqrt{\alpha^2\lambda+1}}))}\right|^2.    
\end{split}
\end{equation}

\subsection{$T\bar{T}$ deformation and BTZ black hole}\label{sec:CS-BTZ}

Following the same procedure in section \ref{sec:CS-AdS3}, one can describe the BTZ blak hole in the formalism of the Chern-Simons theory. In this subsection, we compute the stress–energy tensor of the $T\bar{T}$ deformed boundary theory of the BTZ background. For simplicity, we will focus on the classical solution of the BTZ Chern-Simons theory. We also compare the associated stress–energy tensor with the ``boundary stress–energy tensor'' of the BTZ gravity with a finite cutoff.

The BTZ black hole is described by the metric
\begin{equation}
\begin{split}
&ds^2=-f^2(r)dt^2+f^{-2}(r)dr^2+r^2(d\theta-\omega(r)dt)^2,\\
&f^2(r)=r^2-8GM+\frac{16G^2J^2}{r^2},~~~\omega(r)=\frac{4GJ}{r^2}.
\end{split}
\end{equation}

To describe the BTZ black hole in the Chern-Simons formulation, it is convenient to define
\begin{equation}
J=\frac{b^2-\bar b^2}{4G},~~~M=\frac{-b^2-\bar b^2}{4G},~~~
r=\sqrt{\frac{(1-z^2b^2)(1-z^2\bar b^2)}{z^2}}.
\end{equation}
Then the metric can be written as
\begin{equation}
\label{eq:BTZ-metric-2}
ds^2=\frac{dz^2}{z^2}+\frac{1}{z^2}\left((1-{\bar b^2}{z^2})d\theta+(1+{\bar b^2}{z^2})dt\right)
\left((1-{ b^2}{z^2})d\theta-(1+{ b^2}{z^2})dt\right).
\end{equation}
The associated classical gauge fields are
\begin{equation}
A^{(0)}=\left(\begin{array}{cc}
\frac{dz}{2z}&-zb^2(d\theta+dt)\\ 
z^{-1}(d\theta+dt)&-\frac{dz}{2z}
\end{array} \right),~~~
\bar A^{(0)}=\left(\begin{array}{cc}
\frac{-dz}{2z}&-z^{-1}(d\theta-dt)\\ 
z\bar b^2(d\theta-dt)&\frac{dz}{2z}
\end{array} \right),
\end{equation}
and the group elements are
\begin{equation}
g^{(0)}=\left(
\begin{array}{cc}
\frac{\cos (b x^+ )}{\sqrt{b} \sqrt{z}} & -\sqrt{b} \sqrt{z} \sin (b x^+ ) \\
\frac{\sin (b x^+ )}{\sqrt{b} \sqrt{z}} & \sqrt{b} \sqrt{z} \cos (b x^+ ) \\
\end{array}
\right),~~~
\bar g^{(0)}=\left(
\begin{array}{cc}
\sqrt{\bar b } \sqrt{z} \cos (\bar b  x^-) & -\frac{\sin (\bar b  x^-)}{\sqrt{\bar b } \sqrt{z}} \\
\sqrt{\bar b } \sqrt{z} \sin (\bar b  x^-) & \frac{\cos (\bar b x^-)}{\sqrt{\bar b } \sqrt{z}} \\
\end{array}
\right),
\end{equation}
where $A^{(0)}=(g^{(0)})^{-1} dg^{(0)}$ and $\bar A^{(0)}=(\bar g^{(0)})^{-1} d \bar g^{(0)}$.
The BTZ metric leads to same boundary condition (\ref{eq:bdy-cond}) of gauge fields at boundary. In the same way as in section \ref{sec:CS-AdS3}, we could derive the boundary action and the constrains of the BTZ black hole, which have the same form as (\ref{eq:ac-WZW}) and (\ref{eq:bdy-cond-gauss}) respectively.
However, instead of (\ref{eq:F-phi}), the fields in the BTZ blacck hole are
\begin{equation}
F=\tan (b(\theta+t)),~~~\bar F=\tan (\bar b(\theta-t)),
\end{equation}
which provides the orbit ${\rm Diff}(S^1)/U(1)$.
To describe an BTZ black hole, we require $b^2<0$ and $\bar b^2<0$.
When $b=\bar b\in (0,1/2)$ we have a  conical defect rather than a BTZ black hole. See \cite{Cotler:2018zff} for more discussions.

In Appendix \ref{B}, we show that the solutions to the EOM of the deformed theory can be obtained from the ones of original theory.
The deformed solution associated with $g^{(0)}$ and $\bar g^{(0)}$ is
\begin{equation}
g=g^{(0)}(\tilde x^+)|_{b\rightarrow b_\lambda},~~~
\bar g=\bar g^{(0)}(\tilde x^-)|_{\bar b\rightarrow \bar b_\lambda},
\end{equation}
where
\begin{equation}
\begin{split}
\tilde x^+=&\frac{x^++\lambda\bar b_\lambda^2 x^-}{1-b_\lambda^2 \bar b_\lambda^2 \lambda^2},~~~
\tilde x^-=\frac{x^-+\lambda b_\lambda^2 x^+}{1-b_\lambda^2 \bar b_\lambda^2 \lambda^2},\\
b_\lambda=&\frac{\sqrt{\lambda ^2 \left(b^2-\bar b ^2\right)^2+2 \lambda  \left(b^2+\bar b ^2\right)+1}+b^2 \lambda -\bar b ^2 \lambda -1}{2 b \lambda },\\
\bar b_\lambda=&
\frac{\sqrt{\lambda ^2 \left(b^2-\bar b ^2\right)^2+2 \lambda  \left(b^2+\bar b ^2\right)+1}-b^2 \lambda+\bar b ^2 \lambda -1}{2 \bar b  \lambda }.
\end{split}
\end{equation}
Here we introduce $b_\lambda$ and $\bar b_\lambda$ such that the boundary condition
\begin{equation}
\arctan F |^{\theta=2\pi}_{\theta=0}=2\pi b,~~~
\arctan \bar F |^{\theta=2\pi}_{\theta=0}=2\pi \bar b,
\end{equation}
are undeformed.
The deformed stress–energy tensor  in terms of the classical solution  is
\begin{equation}\label{sttt}
\begin{split}
(T_\lambda)^{\mu}_{~\nu}=&
\frac{k}{\pi}\frac{1}{2-2  b_\lambda ^2 \bar b_\lambda ^2 \lambda ^2}
\left(
\begin{array}{cc}
-2 \lambda  \bar b_\lambda ^2  b_\lambda ^2- b_\lambda ^2-\bar b_\lambda ^2 & \bar b_\lambda ^2- b_\lambda ^2 \\
b_\lambda ^2-\bar b_\lambda ^2 & -2 \lambda\bar b_\lambda ^2    b_\lambda ^2+ b_\lambda ^2+\bar b_\lambda ^2 \\
\end{array}
\right),\\
=&
\frac{1}{8\pi G}\left(
\begin{array}{cc}
\frac{(1-\mathcal{S})}{\lambda} & \bar b^2- b^2 \\
b^2-\bar b^2 &\frac{1}{\lambda} \left(1+\mathcal{S}-\frac{ \mathcal{S}}{\lambda( b-\bar b)^2+1}-\frac{S}{\lambda( b+\bar b)^2+1}\right) \\
\end{array}
\right),
\end{split}
\end{equation}
where
\begin{equation}
\mathcal{S}=\sqrt{1+2 \lambda  \left(b^2+\bar b ^2\right)+\lambda ^2 \left(b^2-\bar b ^2\right)^2}.
\end{equation}

In the following of this subsection, we compare (\ref{sttt}) with the boundary stress–energy tensor of the BTZ black hole  at a cutoff surface $r=r_c$
\cite{Brown:1994gs} in our convention.
Here we mainly follow the derivation in \cite{Kraus:2018xrn}. The boundary stress–energy tensor is define as
\begin{equation}
T_{ij}=\frac{1}{4G}(K_{ij}-K g_{ij}+g_{ij}),
\end{equation}
where $g_{ij}$ is the boundary metric and
$K_{ij}$  the  extrinsic curvature. On
a surface at a finite radial location 
\begin{equation}
z\rightarrow z_c=\left(\frac{\bar b^2+\bar b^2+r_c^2-\sqrt{r_c^4 + 2 (b^2 + \bar b^2) r_c^2 + (b^2 - \bar b^2)^2}}{2 \bar b^2 \bar b^2}\right)^{\frac{1}{2}},
\end{equation}
 we have
\begin{align}
g_{ij}=&
\left(
\begin{array}{cc}
-\frac{\left( b ^2 z_c^2+1\right) \left(\bar b ^2 z_c^2+1\right)}{z_c^2} & \bar b ^2- b ^2 \\
\bar b ^2- b ^2 & \frac{\left( b ^2 z_c^2-1\right) \left(\bar b ^2 z_c^2-1\right)}{z_c^2} \\
\end{array}
\right),\\
K_{ij}=&-z\partial_z g_{ij}|_{z\rightarrow z_c}=
\left(
\begin{array}{cc}
b ^2 \bar b ^2 z_c^2-\frac{ 1}{z_c^2} & 0 \\
0 & \frac{1}{z_c^2}- b ^2 \bar b ^2 z_c^2 \\
\end{array}
\right),
\end{align}
from which we find
\begin{equation}\label{stg}
T^i_{~j}=-\frac{z_c^2}{4 G-4 z_c^4 b ^2 \bar b ^2 G }\left(
\begin{array}{cc}
2z_c^2 \bar b ^2  b ^2+b ^2+\bar b ^2 & b ^2-\bar b ^2 \\
\bar b ^2-b ^2 & 2 z_c^2  \bar b ^2b ^2-b ^2-\bar b ^2 \\
\end{array}
\right).
\end{equation}
We define basis vectors
\begin{align}
v_0^i=&\left(\frac{z_c \sqrt{\left( b^2 z_c^2-1\right) \left(\bar b^2 z_c^2-1\right)}}{1- b^2 \bar b^2 z_c^4},
\frac{z_c^3 \left( b^2-\bar b^2\right)}{\left(1- b^2 \bar b^2 z_c^4\right) \sqrt{\left( b^2 z_c^2-1\right) \left(\bar b^2 z_c^2-1\right)}}\right),\\
v_1^j=&\left(0,\frac{z_c}{\sqrt{\left( b^2 z_c^2-1\right) \left(\bar b^2 z_c^2-1\right)}}\right),
\end{align}
where $v_0$ is a unit vector  normal to a constant $t$ slice of the boundary and $v_1$ is a unit vector  normal to $v_0$.
In the new basis $\{v_I\}$, the stress–energy tensor becomes
\begin{equation}
T^I_{~J}=\frac{z_c^2}{4 G \left(b^2 z_c^2-1\right) \left(\bar b^2 z_c^2-1\right)}
\left(
\begin{array}{cc}
2  b^2 \bar b^2 z_c^2- b^2-\bar b^2 & \bar b^2- b^2 \\
b^2-\bar b^2 & \frac{2  b^4 \bar b^4 z_c^6-3  b^4 \bar b^2 z_c^4-3  b^2 \bar b^4 z_c^4+6  b^2 \bar b^2 z_c^2- b^2-\bar b^2}{ b^2 \bar b^2 z_c^4-1} \\
\end{array}
\right),
\end{equation}
or
\begin{align}
T^I_{~J}=&\frac{1}{4 G r_c^2}
\left(
\begin{array}{cc}
r_c^2(1-\mathcal{S}_c) & \bar b^2- b^2 \\
b^2-\bar b^2 &r_c^2 \left(-\frac{r_c^2 \mathcal{S}_c}{( b-\bar b)^2+r_c^2}-\frac{r_c^2 S}{( b+\bar b)^2+r_c^2}+1+\mathcal{S}_c\right) \\
\end{array}
\right),\\
\mathcal{S}_c=&r_c^{-2}\sqrt{r_c^4 + 2 (b^2 + \bar b^2) r_c^2 + (b^2 - \bar b^2)^2},
\end{align}
which matches (\ref{sttt}) up to a factor under the identification $r_c^2\lambda=1$.
As in \cite{McGough:2016lol,Kraus:2018xrn}, to compare with the energy obtained on the QFT side one should multiply the energy
by the circumference of the circle $L=2\pi r_c$ to get a dimensionless ``proper energy"
\begin{equation}
\mathcal{E}=\frac{L}{2\pi}\int_0^{2\pi} \sqrt{g_{\theta \theta}}T^0_{~0} d\theta=2\pi r_c^2 T^0_{~0}.
\end{equation}
When $r_c$ is large, we have $\mathcal{E}=2 \pi M+O(r_c^{-1})$.

The same result can be derived in the Chern-Simons formulation.
We assume that the boundary term on a finite cutoff surface has the same form as that at infinity
\begin{equation}
\mathcal{L}_{\mathrm{bdy}}=-\frac{1}{4}\mathrm{Tr}( A_\theta  A_\theta)
-\frac{1}{4}\mathrm{Tr}(\bar A_\theta \bar A_\theta).
\end{equation}
The boundary conditions consistent with the variational principle are
\begin{equation}
A^3_\theta=\bar A^3_\theta=0,~~~
A^+_\theta=z^{-1} ,~~~
A^-_\theta=-z^{-1}.
\end{equation}
To obtain the boundary stress–energy tensor we need to insert back the zweibein.
The zweibein on the cutoff surface are
\begin{equation}
E^+_c=\frac{1}{\sqrt{2}}\left((1-{\bar b^2}{z^2})d\theta+(1+{\bar b^2}{z^2})dt\right),~~~
E^-_c=\frac{1}{\sqrt{2}}\left((1-{ b^2}{z^2})d\theta-(1+{ b^2}{z^2})dt\right).
\end{equation}
Therefore the on-shell boundary term should be interpreted as
\begin{equation}
\begin{split}
\mathcal{L}_{\mathrm{bdy}}=&
-\frac{1}{4}\left(\frac{\sqrt{2}\Ept+z \bar A^{(0)+}_t}{\sqrt{2}\Epa+z \bar A^{(0)+}_\theta}\mathrm{tr}(A^{(0)}_\theta A^{(0)}_\theta)
-\frac{\sqrt{2}\Emt-z A^{(0)-}_t}{\sqrt{2}\Ema-z A^{(0)-}_\theta}\mathrm{tr}(\bar A^{(0)}_\theta \bar A^{(0)}_\theta)\right)\\
=&-\frac{1}{2}\left(\frac{\sqrt{2}\Ept-z^2 \bar b^2}{\sqrt{2}\Epa+z^2\bar b^2}b^2
-\frac{\sqrt{2}\Emt+z^2 b^2}{\sqrt{2}\Ema+z^2 b^2}\bar b^2\right).
\end{split}
\end{equation}
Then we get the boundary  stress–energy tensor,
\begin{equation}
\begin{split}
T^{i}_{~j}
=&-\frac{k}{\pi}\frac{1}{\det E}E^A_j\frac{\partial \mathcal{L}_{\mathrm{bdy}}}{\partial E^A_i} |_{E^\pm\rightarrow E^\pm_c}\\
=&\frac{k}{\pi}\frac{1}{2-2 z^4 b ^2 \bar b ^2  }\left(
\begin{array}{cc}
2z^2 \bar b ^2  b ^2+b ^2+\bar b ^2 & b ^2-\bar b ^2 \\
\bar b ^2-b ^2 & 2 z^2  \bar b ^2b ^2-b ^2-\bar b ^2 \\
\end{array}
\right),
\end{split}
\end{equation}
which equals ($\ref{stg}$) up to a factor.

\section{Conclusions and discussions}\label{sec:conclusion}

In this paper, we have studied the $T\bar T$ deformation of chiral bosons.
In particular, the $T\bar T$ deformation of two chiral bosons of opposite chiralities is equivalent to that of a non-chiral free scalar theory at the Hamiltonian level.
Furthermore, we have obtained the all-order $T\bar T$ deformed Lagrangian of more general theories which contain an arbitrary number of chiral bosons with potentials.
Based on these results, we study the $T\bar{T}$ deformation of the boundary theory in Chern-Simons AdS$_3$ gravity which is a constrained chiral WZW model. We have derived the all-order $T\bar{T}$ deformed Lagrangian and computed the one-loop torus partition function of the deformed theory,
which satisfies the flow equation of general $T \bar T$ torus partition function up to one-loop order. Our result suggests that the one-loop torus partition function is not one-loop exact under the $T\bar{T}$ deformation, which is unlike the situation in the undeformed theory \cite{Cotler:2018zff}. Moreover, we have computed the stress–energy tensor of the solution associated with a BTZ black hole in the deformed theory, which matches the boundary stress–energy tensor of the BTZ black hole at a finite radial location on the bulk side.

Let us comment on future research directions. It would be interesting to start with the Chern-Simons theory describing the AdS$_3$ gravity with a finite cutoff to derive the $T\bar{T}$ deformed boundary action. This will help us to realize the holography under $T\bar{T}$ deformation more explicitly. Moreover, the original exact boundary action of Chern-Simons AdS$_3$ gravity can be applied to compute the four-point functions in the light-light and heavy-light limit \cite{Cotler:2018zff}. Recently, many studies have been devoted to the correlators in general $T\bar T$ deformed CFTs \cite{Cardy:2019qao,He:2019vzf,He:2020udl,He:2019ahx,Li:2020pwa}. It would be interesting to compute correlators in our deformed model and compare them with these results.
It would also be interesting to generalize our analysis to higher spin theories of gravity formulated in terms of $SL(N,\mathbb{R})$ Chern-Simons theory \cite{Ma:2019gxy}.

\subsection*{Acknowledgements}
We would like to thank Sergey Frolov, Song He, Yunfeng Jiang, Matias Leoni, Chen-Te Ma and Pujian Mao for useful discussions and comments on the draft.
The works of H.O and H.S. are supported by the grant ``Exact Results in Gauge and String Theories'' from the Knut and Alice Wallenberg foundation.

\appendix

\section{$J \bar J$  and $T \bar J$ deformation of two chiral bosons }\label{A}

In this appendix, we consider the $J\bar{J}$ and $T\bar{J}$ deformation of the chiral bosons. 

\subsection{$J \bar J$ deformation}
Consider the Lagrangian of wo chiral bosons of opposite chiralities
\begin{equation}
\mathcal{L}_0=\frac{1}{2} \left( \ft\fa- \fa \fa-\gt\ga-\ga \ga \right).
\end{equation}
To define currents $J$ and $\bar J$, we couple the chiral bosons to gauge fields
\begin{equation}
\begin{split}
\mathcal{L}_0=&\frac{1}{2} \left( \ft\fa- \fa \fa
-(A_\theta-A_t)(2\fa+A_\theta)\right)\\
&+\frac{1}{2} \left(-\gt\ga-\ga \ga-(\bar A_\theta+\bar A_t)(2\ga+\bar A_\theta) \right).
\end{split}
\end{equation}
We define the currents as
\begin{equation}
J^i=\frac{\partial{\mathcal{L}}}{\partial A^i},
~~~\bar J^i=\frac{\partial{\mathcal{L}}}{\partial \bar A^i}.
\end{equation}
In the undeformed theory
\begin{equation}
\partial_i J_0^i=\frac{1}{2}\partial_t A_\theta
-\frac{1}{2}\partial_\theta A_t,~~~
\partial_i \bar J_0^i=\frac{1}{2}\partial_\theta \bar A_t
-\frac{1}{2}\partial_t \bar A_\theta.
\end{equation}
When $A$ and $\bar A$ are closed, $J$ and $\bar J$ are conserved. The $J\bar J$ operator in the deformed theory is defined as
\begin{equation}
(J\bar J)_\lambda=2J_\lambda^t \bar J^\theta_\lambda-2\bar J^t_\lambda J^\theta_\lambda.
\end{equation}
Solving the flow equation
\begin{equation}
\frac{\partial \mathcal{L}_\lambda}{\partial \lambda}
=(J\bar J)_\lambda,
\end{equation}
We get
\begin{equation}
\begin{split}
\mathcal{L}_\lambda=&\mathcal{L}_0-\frac{4\lambda(\fa^2\lambda+\ga^2\lambda+\fa\ga(1+\lambda^2))}{(1-\lambda^2)^2}
-\frac{\lambda }{\lambda ^2-1}(\At \ga-\Bt \fa+\frac{\At \Bx-\Ax \Bt}{2})\\
&-\frac{\lambda^2 }{\lambda ^2-1}(\At \fa-\Bt \ga+\frac{\At \Ax-\Bx \Bt}{2})-\frac{2\lambda }{(\lambda ^2-1)^2}\Ax \Bx
+\frac{\lambda ^2 \left(\lambda ^2-3\right)}{2 \left(\lambda ^2-1\right)^2}(\Ax^2+\Bx^2)\\
&+\frac{\lambda ^2 \left(\lambda ^2-5\right)}{\left(\lambda ^2-1\right)^2}(\Ax \fa+\Bx \ga)
-\frac{\lambda  \left(\lambda ^2+3\right)}{\left(\lambda ^2-1\right)^2}(\Bx \fa+\Ax \ga).
\end{split}
\end{equation}
Finally setting $A=\bar A=0$, we get
\begin{equation}
\begin{split}
\mathcal{L}_\lambda=&\mathcal{L}_0-\frac{4\lambda(\fa^2\lambda+\ga^2\lambda+\fa\ga(1+\lambda^2))}{(1-\lambda^2)^2}\\
=&\mathcal{L}_0-4\lambda \fa \ga-4\lambda^2(\fa^2+\ga^2)+O\left(\lambda ^3\right).
\end{split}
\end{equation}
\subsection{$T \bar J$ deformation}
We couple the left chiral boson to the zweibein and the left chiral boson to a gauge field
\begin{equation}
\mathcal{L}_0=-\frac{1}{2} \left( -\ft\fa+\frac{\Ept}{\Epa} \fa \fa \right)
+\frac{1}{2} \left(-\gt\ga-\ga \ga-(\bar A_\theta+\bar A_t)(2\ga+\bar A_\theta) \right).
\end{equation}
We define the currents as
\begin{equation}
T_+^i=\frac{\partial{\mathcal{L}}}{\partial E^+_i},
~~~\bar J^i=\frac{\partial{\mathcal{L}}}{\partial \bar A^i}.
\end{equation}
 The $T\bar J$ operator in the deformed theory is defined as
 \begin{equation}
(T\bar J)_\lambda=2T_{+\lambda}^t \bar J^\theta_\lambda-2\bar J^t_\lambda T^\theta_{+\lambda} ,
\end{equation}
Solving the flow equation
\begin{equation}
\frac{\partial \mathcal{L}_\lambda}{\partial \lambda}
=(T\bar J)_\lambda,
\end{equation}
We get
\begin{equation}
\begin{split}
\mathcal{L}_\lambda=&-\frac{1}{2} \left( -\ft\fa+\fa \fa \right)
+\frac{1}{2} \left(-\gt\ga-\ga \ga-(\bar A_\theta+\bar A_t)(2\ga+\bar A_\theta) \right)\\
&+\frac{\Epa (2 (\Epa+\Ept)-\lambda  (\bar A_\theta+\bar A_t))}{2 \lambda ^2}\\
&\times\left( \sqrt{1+\frac{\lambda}{\Epa}(\bar A_\theta+2\ga)+\frac{\lambda ^2}{\Epa {}^2}((\ga+\frac{\bar A_\theta}{2})^2)-\fa^2}-1-
\frac{\lambda}{\Epa}(\ga+\frac{\bar A_\theta}{2})\right).
\end{split}
\end{equation}

Finally we set $\Epa-1=\Ept-1=\bar A=0$ and obtain
\begin{equation}
\begin{split}
\mathcal{L}_\lambda=&\frac{1}{2} \left( \ft\fa- \fa \fa-\gt\ga-\ga \ga \right)\\
&+\frac{2 \left(\sqrt{\lambda ^2 \left(\ga^2-\fa^2\right)+2 \ga\lambda +1}-\ga \lambda -1\right)}{\lambda ^2}.
\end{split}
\end{equation}

\section{$T \bar T$ deformed  chiral WZW model}\label{B}
We consider the sum of a left and a right chiral WZW model
\begin{equation}
S=S_-[g]+S_+[\bar g],
\end{equation}
with 
\begin{equation}
 S_{\pm}[g]=\frac{k}{2\pi}\left( \int d^2x \,\text{Tr} \big((g^{-1})'\partial_{\pm} g\big) 
  \mp \frac{1}{6}\int_{B}\text{Tr}(g^{-1}dg \wedge g^{-1}dg \wedge g^{-1}dg)\right),
\end{equation}
where $g$ and $\bar g$ are group elements of group $G$ and $\bar G$ respectively. 
We define
\begin{equation}
A_i=g^{-1}\partial_i g,~~~
\bar A_i=\bar g^{-1}\partial_i \bar g.
\end{equation}
The equations of motions are
\begin{equation}
\partial_-A_\theta=\partial_+\bar A_\theta=0.
\end{equation}
Consider the $T \bar T$ deformed  chiral WZW model 
\begin{equation}
\begin{split}
\frac{2\pi}{k}S_\lambda=&
\frac{1}{2} \int d^2x \Big((\mathrm{tr}( A_\theta A_t)-\mathrm{tr}( \bar A_\theta  \bar A_t) +\frac{1}{{\lambda}}(\cS-1) \Big)\\
&+ \frac{1}{6}\int_{B}\Big(\mathrm{tr}(g^{-1}dg \wedge g^{-1}dg \wedge g^{-1}dg)
-\mathrm{tr}(\bar g^{-1}d \bar g \wedge \bar g^{-1}d \bar g \wedge \bar g^{-1}d\bar g) \Big),
\end{split}
\end{equation}
with
\begin{equation}
\cS=\sqrt{1
	-2{(\mathrm{tr}( A_\theta A_\theta)+\mathrm{tr}( \bar A_\theta \bar A_\theta))}\lambda
	+(\mathrm{tr}( A_\theta A_\theta)-\mathrm{tr}( \bar A_\theta \bar A_\theta))^2\lambda^2}.
\end{equation}
The equations of motions are
\begin{align}
&\partial_\theta \Big( \frac{1-\lambda(\mathrm{tr}( A_\theta A_\theta)-\mathrm{tr}( \bar A_\theta \bar A_\theta))}{\cS} A_\theta \Big)
-\partial_t A_\theta=0,\label{e1wzw}\\
&\partial_\theta \Big( \frac{1+\lambda(\mathrm{tr}( A_\theta A_\theta)-\mathrm{tr}( \bar A_\theta \bar A_\theta))}{\cS} \bar A_\theta \Big)
+\partial_t \bar A_\theta=0 \label{e2wzw}.
\end{align}
When $\lambda=0$, a solution to the  equation of motion is
\begin{equation}
g=h(t) g_0(x^+),~~~\bar g=\bar h(t) \bar g_0(x^-).
\end{equation}
We introduce a new set of coordinate $(\tilde t,\tilde \theta)$ and define a field dependent coordinate
transformation with the Jacobian
\begin{equation}
\begin{split}
\left(\begin{array}{cc}
\partial_{\tilde t} t&\partial_{\tilde t} \theta \\ 
\partial_{\tilde \theta} t&\partial_{\tilde\theta} \theta
\end{array} \right)
&=\left(
\begin{array}{cc}
-\frac{F_{\tilde \theta \tilde \theta} \lambda }{2}-\frac{\bar F_{\tilde \theta \tilde \theta} \lambda }{2}+1 & \frac{F_{\tilde \theta \tilde \theta} \lambda }{2}-\frac{\bar F_{\tilde \theta \tilde \theta} \lambda }{2} \\
\frac{\bar F_{\tilde \theta \tilde \theta} \lambda }{2}-\frac{F_{\tilde \theta \tilde \theta} \lambda }{2} & \frac{F_{\tilde \theta \tilde \theta} \lambda }{2}+\frac{\bar F_{\tilde \theta \tilde \theta} \lambda }{2}+1 \\
\end{array}
\right),
\end{split}
\end{equation}
where 
\begin{align}
F_{\tilde \theta \tilde \theta}&=\mathrm{tr}(g_0^{-1}(\tilde x^+) \partial_{\tilde\theta} g_0(\tilde x^+) 
g_0^{-1}(\tilde x^+)  \partial_{\tilde\theta}g_0(\tilde x^+) ),~~~\\
\bar F_{\tilde \theta \tilde \theta}&=\mathrm{tr}(\bar g_0^{-1}(\tilde x^-) \partial_{\tilde\theta} \bar g_0(\tilde x^-) 
\bar g_0^{-1}(\tilde x^-)  \partial_{\tilde\theta}\bar g_0(\tilde x^-) ),\\
\tilde{x}^\pm&= \tilde\theta \pm \tilde t.
\end{align}

Then the solution
\begin{equation}
g(t,\theta)=h(t) g_0(\tilde x^+(t,\theta)),~~~\bar g(t,\theta)=\bar h(t) \bar g_0(\tilde x^-(t,\theta)).
\end{equation}
satisfies the equation of motion for the deformed theory.
Using
\begin{align}
&\partial_\theta+\partial_t=2\frac{\partial_{\tilde x^+}-\lambda F_{\tilde \theta \tilde \theta}\partial_{\tilde x^-}}{1-\lambda^2 F_{\tilde \theta \tilde \theta} \bar F_{\tilde \theta \tilde \theta}},\\
&\partial_\theta-\partial_t=2\frac{\partial_{\tilde x^-}-\lambda \bar F_{\tilde \theta \tilde \theta}\partial_{\tilde x^+}}{1-\lambda^2 F_{\tilde \theta \tilde \theta} \bar F_{\tilde \theta \tilde \theta}},\\
&A_\theta=g_0^{-1}(\tilde x^+) \partial_{\tilde\theta} g_0(\tilde x^+) \frac{1-\lambda\bar F_{\tilde \theta \tilde \theta}}{1-\lambda^2 F_{\tilde \theta \tilde \theta} \bar F_{\tilde \theta \tilde \theta}},\\
&\bar A_\theta=\bar g_0^{-1}(\tilde x^-) \partial_{\tilde\theta} \bar g_0(\tilde x^-) \frac{1- \lambda F_{\tilde \theta \tilde \theta}}{1-\lambda^2 F_{\tilde \theta \tilde \theta} \bar F_{\tilde \theta \tilde \theta}},\\
&\frac{1-\lambda(\mathrm{tr}( A_\theta A_\theta)-\mathrm{tr}( \bar A_\theta \bar A_\theta))}{\cS} A_\theta
=g_0^{-1}(\tilde x^+) \partial_{\tilde\theta} g_0(\tilde x^+) \frac{1+\lambda\bar F_{\tilde \theta \tilde \theta}}{1-\lambda^2 F_{\tilde \theta \tilde \theta} \bar F_{\tilde \theta \tilde \theta}},\\
&\frac{1-\lambda(\mathrm{tr}( A_\theta A_\theta)-\mathrm{tr}( \bar A_\theta \bar A_\theta))}{\cS} A_\theta
=\bar g_0^{-1}(\tilde x^-) \partial_{\tilde\theta} \bar g_0(\tilde x^-) \frac{1+ \lambda F_{\tilde \theta \tilde \theta}}{1-\lambda^2 F_{\tilde \theta \tilde \theta} \bar F_{\tilde \theta \tilde \theta}},
\end{align}
one can check that (\ref{e1wzw}) and  (\ref{e2wzw}) are satisfied.
One should also consider boundary condition so in general $g_0$ and $\bar g_0$ should depend on $\lambda$.

%\bibliographystyle{JHEP}  
%\bibliography{ref}

\providecommand{\href}[2]{#2}\begingroup\raggedright\endgroup

\end{document}